\documentclass[aip,apl,reprint,twocolumn,superscriptaddress,amsmath,amssymb]{revtex4}

\usepackage[]{graphicx}
\usepackage[utf8]{inputenc}
\usepackage[T1]{fontenc}%
\usepackage{dcolumn}%
\usepackage{bm}%
\usepackage{lmodern}
\usepackage{hyperref}
\usepackage{xcolor}
\PassOptionsToPackage{hyphens}{url}

\newcommand{\beq}{\begin{equation}\begin{aligned}}
\newcommand{\eeq}{\end{aligned}\end{equation}}

\newcommand{\mum}{$\mu$m}
\newcommand{\ws}{WS\ensuremath{_2}}

\definecolor{linkcol}{rgb}{0,0,0.4}
\definecolor{citecol}{rgb}{0.5,0,0}

\hypersetup{
bookmarksopen=true,
pdftitle="Scanning photocurrent microscopy reveals electron-hole asymmetry in ionic liquid-gated \ws{} transistors",
pdfauthor="N. Ubrig et al.",
pdfsubject="Scanning photocurrent microscopy reveals electron-hole asymmetry in ionic liquid-gated \ws{} transistors", 
pdfstartview={FitH},    
pdfmenubar=true, 
pdfhighlight=/O, 
colorlinks=true, 
pdfpagemode=UseNone, 
pdfpagelayout=SinglePage, 
pdffitwindow=true, 
linkcolor=linkcol, 
citecolor=linkcol, 
urlcolor=blue
}

\begin{document}


\title{Scanning photocurrent microscopy reveals electron-hole asymmetry in ionic liquid-gated \ws{} transistors}
\author{Nicolas Ubrig}
\email{nicolas.ubrig@unige.ch}
\affiliation{DPMC, Université de Genève, 24 quai Ernest Ansermet, CH-1211, Geneva, Switzerland}
\author{Sanghyun Jo}
\affiliation{DPMC, Université de Genève, 24 quai Ernest Ansermet, CH-1211, Geneva, Switzerland}
\affiliation{GAP, Université de Genève, 24 quai Ernest Ansermet, CH-1211, Geneva, Switzerland}
\author{Helmuth Berger}
\affiliation{Institut de Physique de la Matière Condendée, Ecole Polytechnique Fédérale de Lausanne, CH-1015, Lausanne, Switzerland}
\author{Alberto F. Morpurgo}
\affiliation{DPMC, Université de Genève, 24 quai Ernest Ansermet, CH-1211, Geneva, Switzerland}
\affiliation{GAP, Université de Genève, 24 quai Ernest Ansermet, CH-1211, Geneva, Switzerland}
\author{Alexey B. Kuzmenko}
\email{Alexey.Kuzmenko@unige.ch}
\affiliation{DPMC, Université de Genève, 24 quai Ernest Ansermet, CH-1211, Geneva, Switzerland}

\date{\today}

\begin{abstract}
We perform scanning photocurrent microscopy on \ws{} ionic li\-quid-gated field effect transistors exhibiting high-quality ambipolar transport. By properly biasing the gate electrode we can invert the sign of the photocurrent showing that the minority photocarriers are either electrons or holes. Both in the electron- and the hole-doping regimes the photocurrent decays exponentially as a function of the distance between the illumination spot and the nearest contact, in agreement with a two-terminal Schottky-barrier device model. This allows us to compare the value and the doping dependence of the diffusion length of the minority electrons and holes on a same sample. Interestingly, the diffusion length of the minority carriers is several times larger in the hole accumulation regime than in the electron accumulation regime, pointing out an electron-hole asymmetry in \ws{}.
\end{abstract}


\maketitle

\begin{figure}
 \centering
 \includegraphics[width=.24\textwidth]{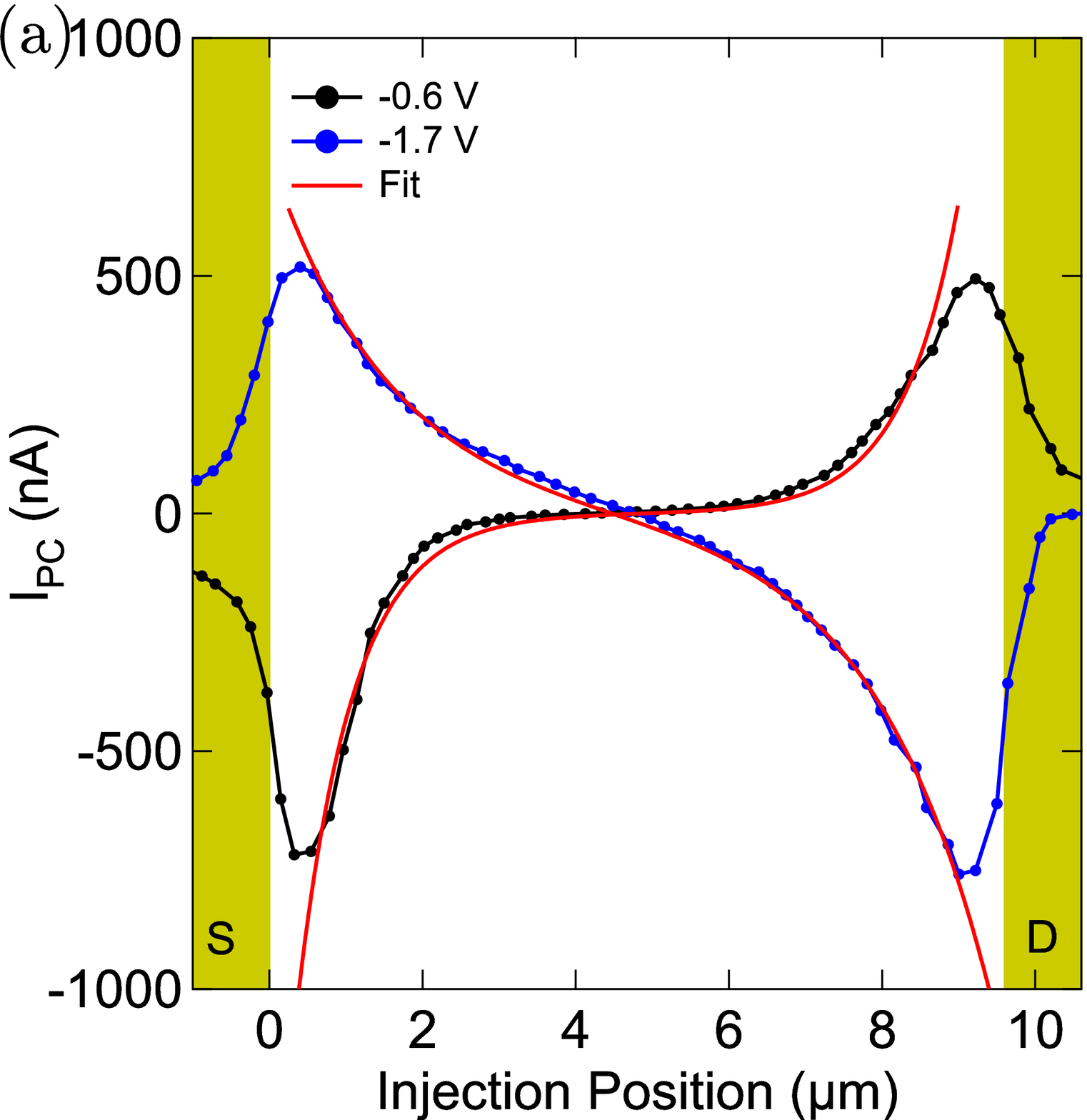}%
 \hfill%
 \includegraphics[width=.24\textwidth]{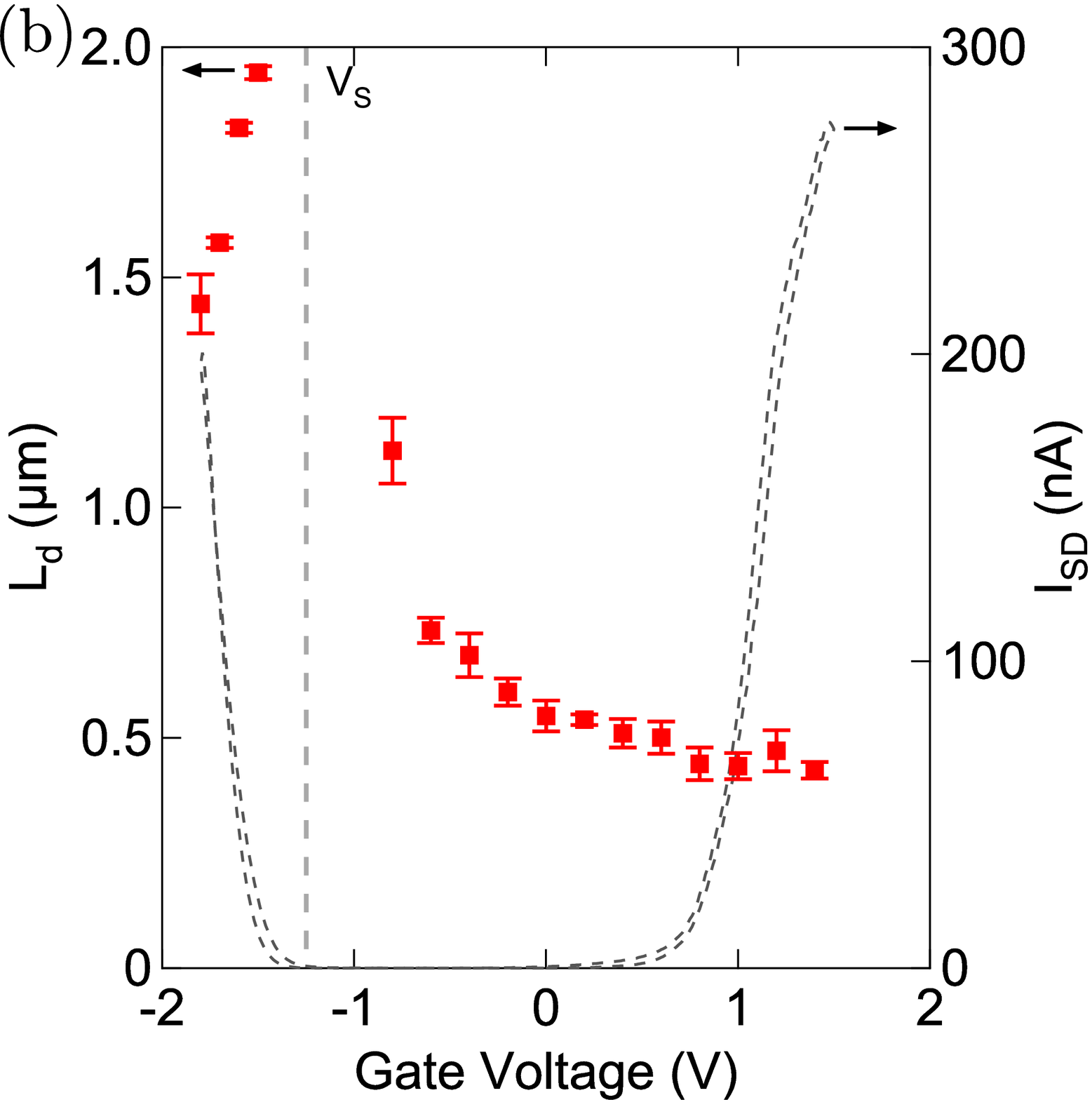}%
 \caption{(a) SPCM line-scans and corresponding fits for $V_G$-0.6 V and -1.7 V. (b) The minority carrier diffusion length as function of gate voltage obtained by fitting SPCM data (red squares). The dashed line is the transport curve for the same sample. The horizontal line indicates $V_S$.}
 \label{fig:difflength}
\end{figure}

Semiconducting transition metal dichalcogenides, MX$_2$ (M = W, Mo, X = S,Se, Te), are long known as efficient photovoltaic materials \cite{tributsch_electrochemical_1979,kam_detailed_1982,decker_photoelectrochemical_1992,gourmelon_ms2_1997,shanmugam_schottky-barrier_2012,britnell_strong_2013} with the band gap covering a broad and technologically important energy range between 0.5 and 2 eV. In the last years, they attracted a lot of interest due to the sensitivity of their band structure to thickness, with a strong dependence of the band gap on the number of layers. A spectacular transition from an indirect gap in multilayer flakes to a direct gap in monolayers was observed in MoS$_2$ \cite{splendiani_emerging_2010,mak_atomically_2010}, WS$_2$ \cite{gutierrez_extraordinary_2012,zhao_evolution_2013,zeng_optical_2013} and WSe$_2$ \cite{zhao_evolution_2013,zeng_optical_2013}, with a dramatic effect on optical properties such as photoluminescence. Another major recent development is the fabrication of field effect transistors (FETs) based on monolayer MoS$_2$ with a high on/off ratio ($\sim 10^8$) demonstrating also the potential of these compounds in electronics\cite{radisavljevic_single-layer_2011}.\\
In this context, it is important, from the point of view of fundamental physics and optoelectronic applications, to explore and to learn to control the dynamics of photo-excited carriers in transition metal dichalcogenides in a broad range of doping regimes. To this end, scanning photocurrent microscopy (SPCM) is particularly useful, as it allows one to probe the electronic transport properties of minority carriers which contribute mostly to the photocurrent. Indeed it was successfully applied to semiconducting nanowires (Si, CdS, PbS), carbon nanotubes\cite{Ahn_Scanning_2005,yang_controlled_2012,Gu_Near-field_2005,Balasubramanian_Photoelectronic_2004} and more recently to graphene\cite{xia_photocurrent_2009} and mono-\cite{buscema_large_2013}/few-layer \cite{wu_elucidating_2013} MoS$_2$. \\
Despite the relatively large value of the band gap in transition metal dichalcogenides, ambipolar operation was realized recently\cite{podzorov_high-mobility_2004,sik_hwang_transistors_2012} using conventional gating techniques. However, the use of ionic liquid gates generally results in much higher carrier densities than using a solid-state dielectric \cite{panzer_low-voltage_2005,shimotani_electrolyte-gated_2006,ueno_electric-field-induced_2008,braga_quantitative_2012,ye_superconducting_2012}. Ambipolar ionic liquid gated-FET based on transition metal dichalcogenides\cite{braga_quantitative_2012,zhang_ambipolar_2012} showed almost ideal transport characteristics at low operating gate voltage. So far, however, ionic liquid gating of transition metal dichalcogenides has not been  employed on optical experiments.\\
In this paper, we apply scanning photocurrent microscopy to ionic liquid gated FET devices similar to those studied in Ref. [\onlinecite{braga_quantitative_2012}]. This allows us to measure the dependence of the photocurrent on the light injection position, which we explain with help of a simple 1D band model of the device. From the analysis of the SPCM curves we extract the diffusion length of the minority carriers and find different photocurrent decay lengths for electron and hole accumulation, pointing out a strong electron-hole asymmetry in \ws{}.\\
\begin{figure}
 \centering
\includegraphics[width=.2\textwidth]{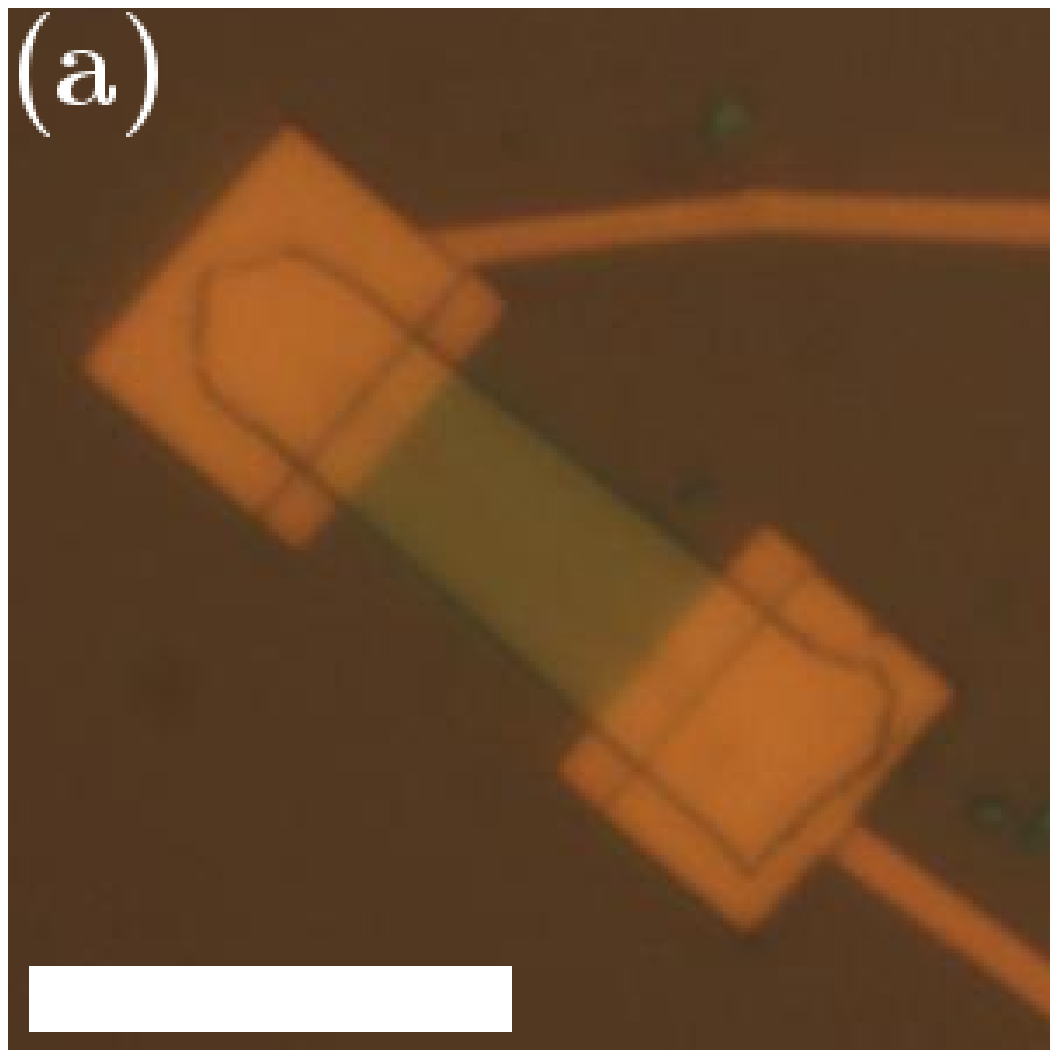}%
\hspace{0.02\textwidth}%
\includegraphics[width=.2\textwidth]{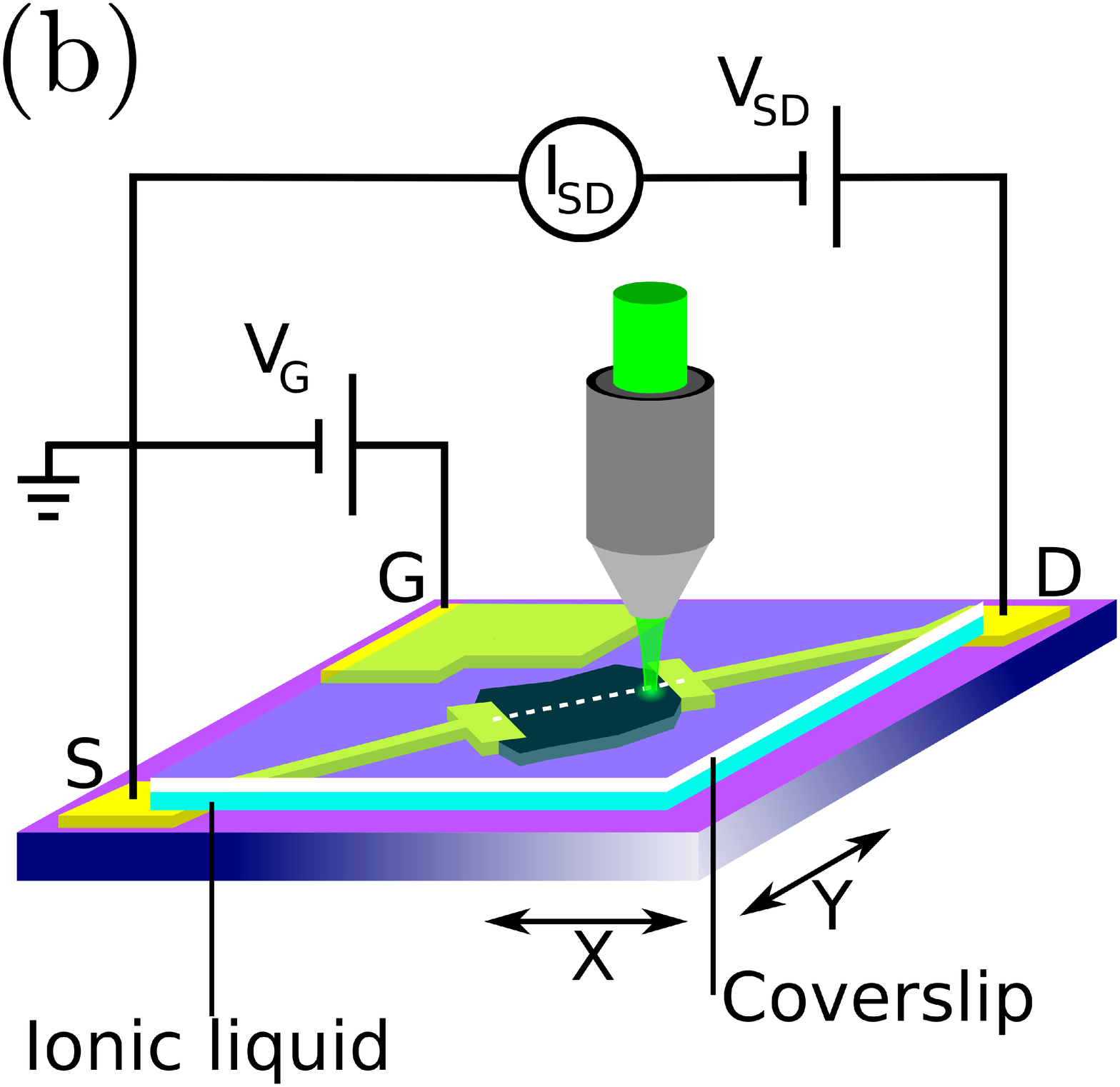}%
\caption{(a) Optical image of a two-terminal WS$_2$ device immersed into ionic-liquid. The scale bar corresponds to 10 \mum{}. (b) Schematic representation of the device and the experimental setup.}
\label{fig:sampledrawing}
\end{figure}
Mechanically exfoliated \ws{} flakes were transfered on a Si/SiO$_2$ substrate. Their typical thickness was 60-80 nm as determined by the atomic force microscopy (AFM). The devices were fabricated through electron-beam lithography, evaporation of Ti/Au contacts and subsequent lift-off. The region under the contacts was bombarded with Ar-ions prior to metal deposition. A droplet of transparent ionic liquid (\emph{1-ethyl-3-methylimidazolium-bis(trifluoromethylsulfonyl)imide}, commonly referred to as EMI-TFSI), was placed onto the substrate in a way to cover the flake and a gate electrode. A glass coverslip was put on top, which allowed us to obtain sharp optical images and ensure a good laser focusing on the sample during the SPCM experiments. A typical image of a contacted sample under ionic liquid is shown in Figure \ref{fig:sampledrawing}(a). Immediately after the fabrication, the devices were introduced into an optical cryostat, where they were kept at low pressure ($ < 10^{-6}$ mbar) for at least 24 hours before the experiments, in order to remove moisture from the ionic liquid.\\
The experimental setup combining electrical transport characterization and photocurrent microscopy is schematically shown in Figure \ref{fig:sampledrawing}(b). Two electrical contacts on the sample are used as source ({\sf S}) and drain ({\sf D}), while a large metal region close to the sample playing the role of the gate electrode ({\sf G}). The gate area is made intentionally much larger than those of the sample and the contacts in order to maximize the gate efficiency. All measurements were performed at room temperature and the gate voltage was limited to $\pm$1.8 V to avoid any chemical reactions at the WS$_2$/liquid interface. We prepared and studied several devices, showing similar transport behavior and SPCM patterns. In this paper we discuss only effects reproducibly observed in all the investigated samples.\\
\begin{figure}
   \centering%
	\includegraphics[width=.235\textwidth]{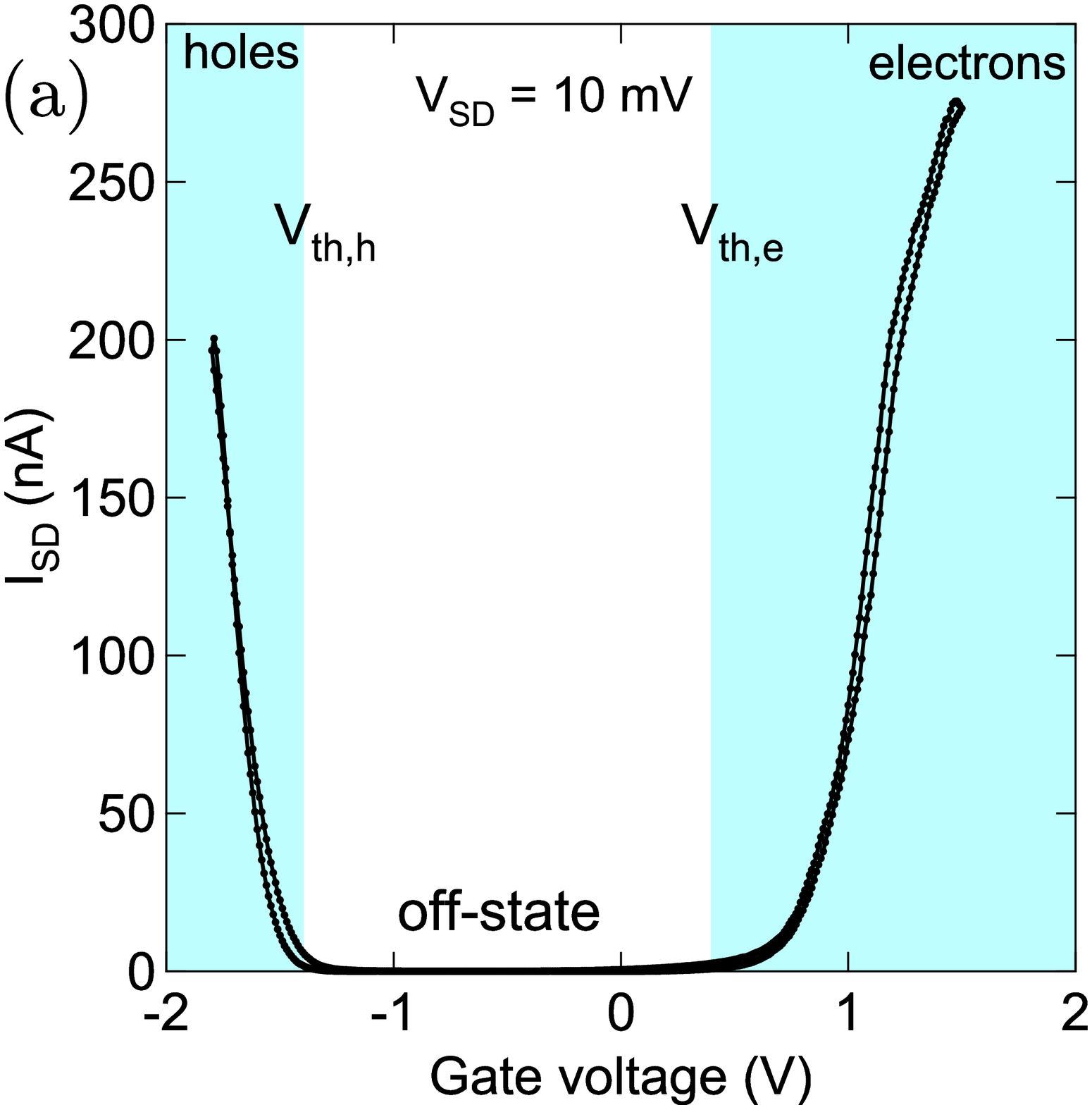} %
	\hfill%
    \includegraphics[width=.235\textwidth]{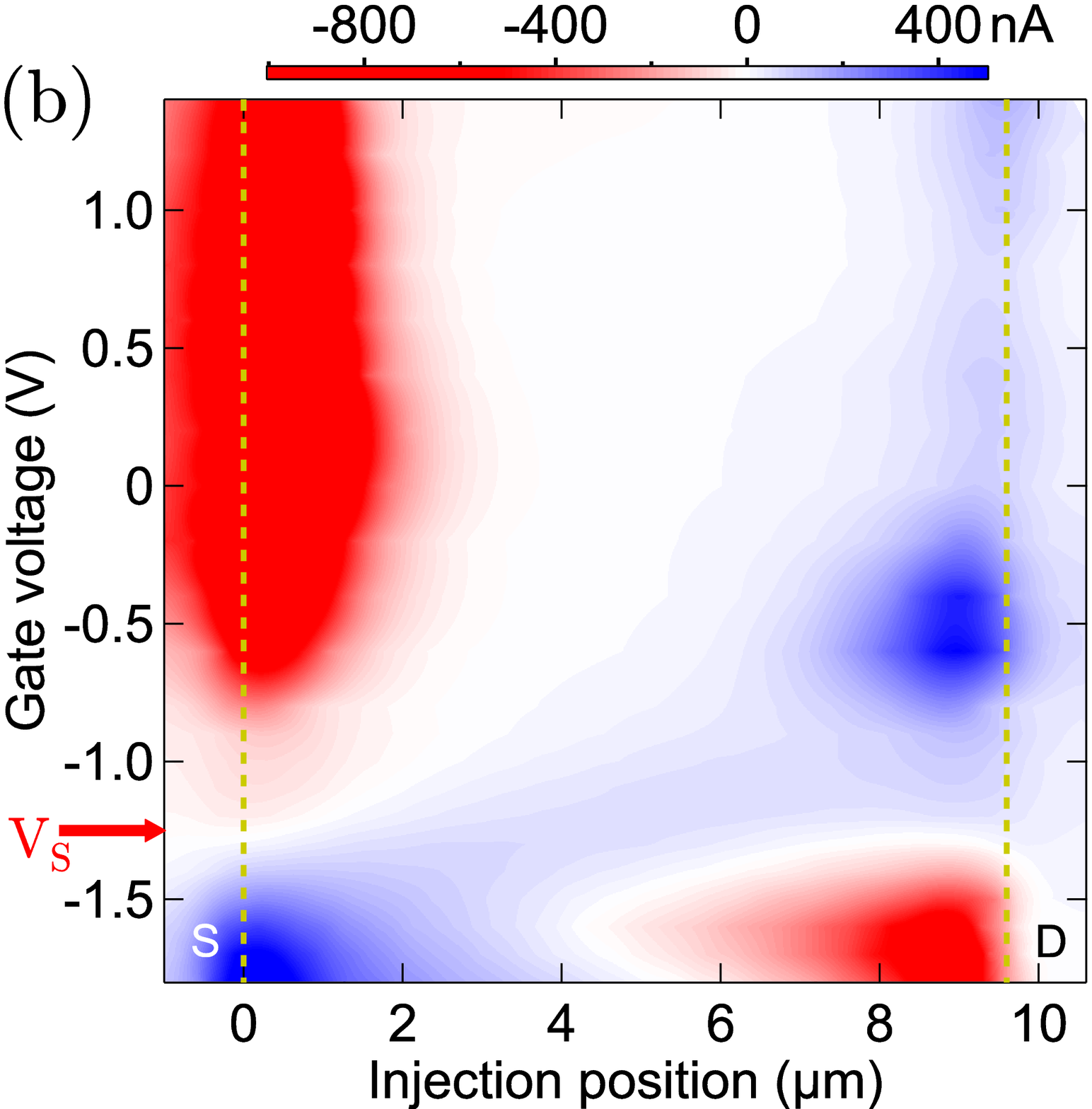}%
 \caption{(a) Transfer characteristic at dark for the device shown in Fig. \ref{fig:sampledrawing}(a) at V$_{\rm SD}$ = 10 mV. The blue areas indicate the regimes with the hole and electron transport. (b) False-color map of the photocurrent as a function of the gate voltage at V$_{\rm SD}$ = 0 V and the laser spot position. The yellow lines indicate the contact edge. The flat band voltage $V_S\approx$ -1.25 V is indicated.}
 \label{fig:PCcolor}
\end{figure}%
Figure \ref{fig:PCcolor}(a) presents the transfer characteristic without illumination of the device shown in Figure \ref{fig:sampledrawing}(a). It shows the dependence of the source-drain current, $I_{SD}$, on the gate voltage, $V_G$, measured at a bias voltage V$_{\rm SD}$ = 10 mV. We observe excellent ambipolar transport behavior, with high on-state current in both the electron- and hole regime, negligibly small off-state current, and hysteresis-free behavior. During this measurement, the gate leakage current was negligibly small. These observations demonstrate the high quality of the devices. The electron and hole threshold voltages are respectively $V_{\rm th,e}=$+0.4 V and $V_{\rm th,h}=$-1.4 V. The difference between these values (1.8 V) is somewhat larger than the band gap of WS$_2$ ($\Delta/e \sim $1.4 V), which is likely due to a finite potential difference between the gate electrode and the liquid (in these devices no reference electrode is present). \cite{braga_quantitative_2012}.\\
The SPCM measurements were performed using an Ar-laser with the wavelength $\lambda = 514.5$ nm (green) at a power of $P$=60 $\mu$W. We verified that for such a low laser power the photocurrent depends linearly on $P$. Light was injected into a microscope objective and focused on the sample surface on a spot of the diameter of about 1 \mum{} (Figure \ref{fig:sampledrawing}(b)). For a given value of $V_G$ and the bias voltage set to zero (V$_{\rm SD}$ = 0 V), the position-dependent photocurrent between the two contacts, $I_{PC}(x)$, was recorded as the sample was scanned in the horizontal plane using an XY-stage so that the laser spot travels  from one contact to another along a straight line in the middle of the sample, as indicated by white dashed line in Figure \ref{fig:sampledrawing}(b). The exact position, $x$, of the laser spot with respect to the contact electrodes was determined with the help of an integrated optical camera.\\
\begin{figure*}
  \centering%
  \includegraphics[width=.33\textwidth]{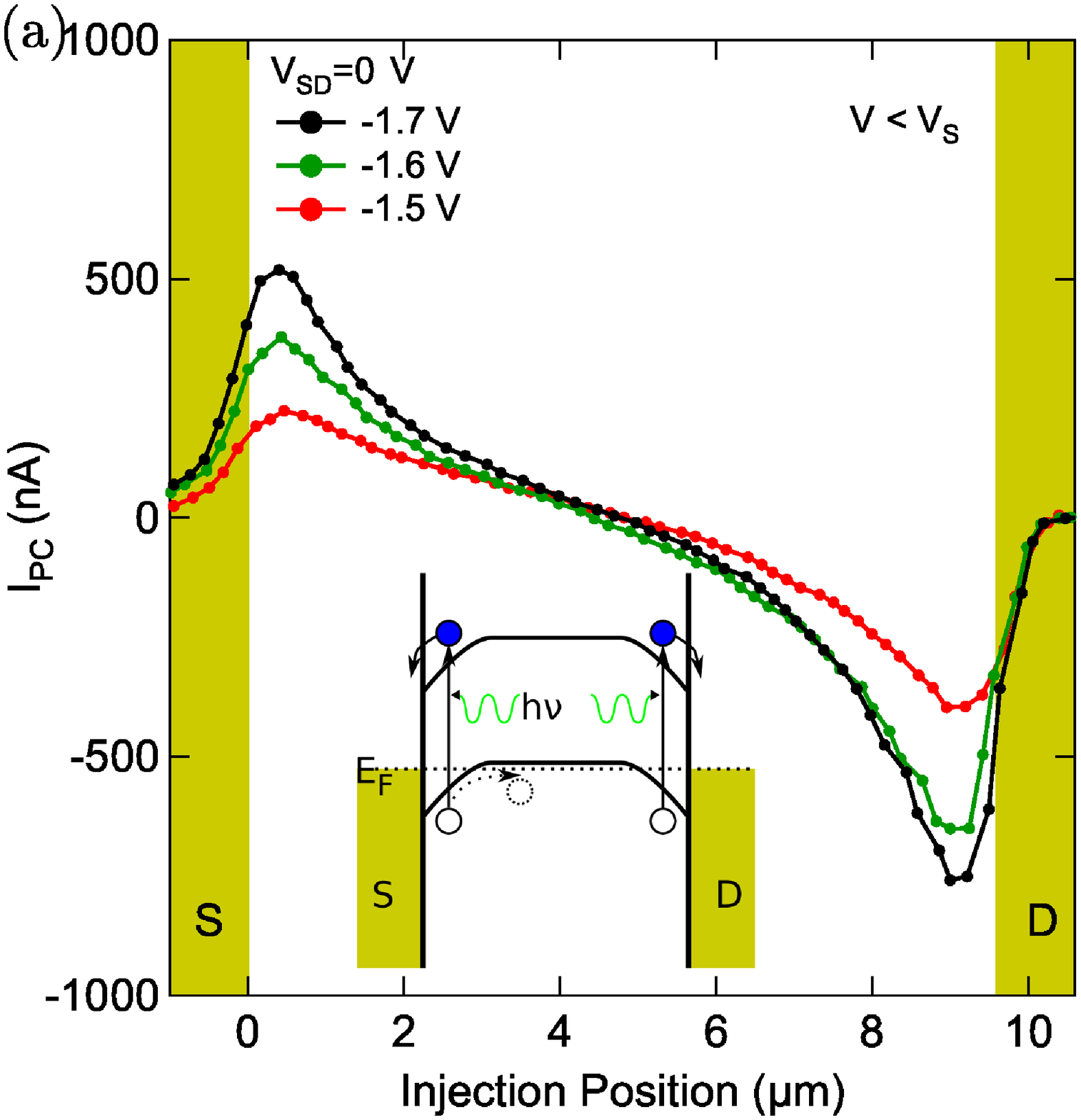}%
  \hfill%
  \includegraphics[width=.33\textwidth]{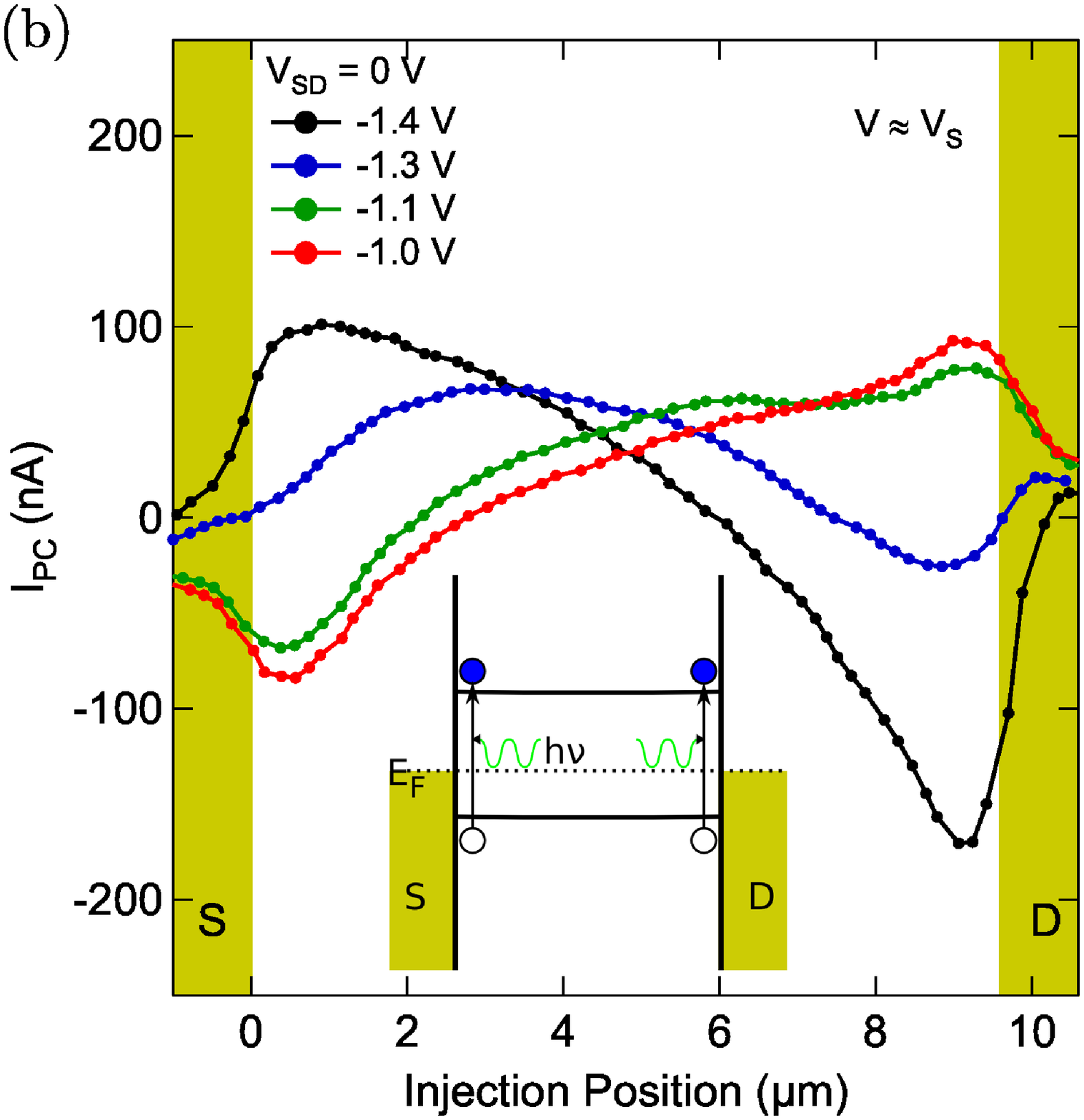}%
  \hfill%
  \includegraphics[width=.33\textwidth]{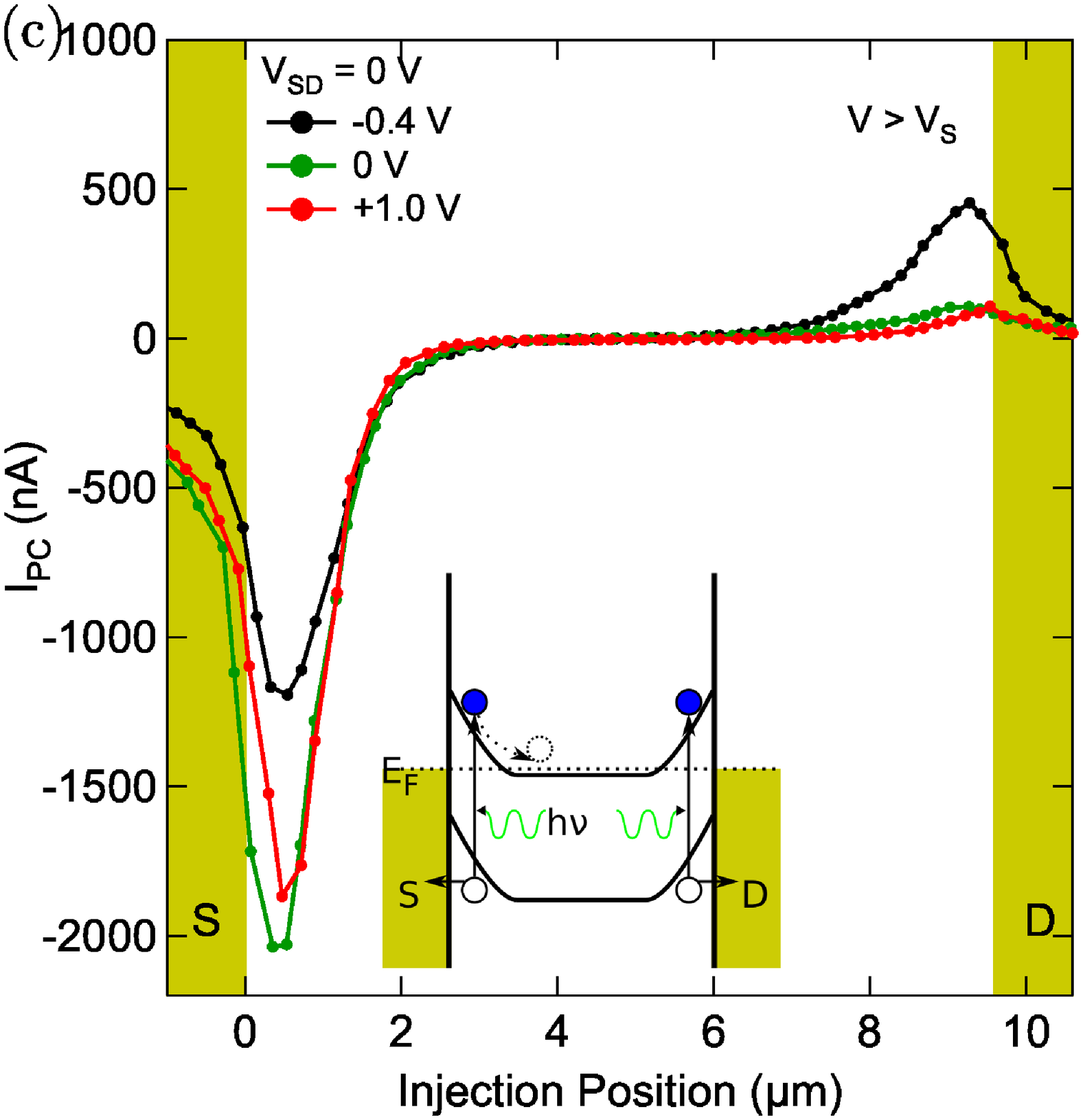}%
\caption{Photocurrent line scans at different values of the gate voltage at V$_{\rm SD}$ = 0 V. Three regimes are shown separately: downward band bending (a), band flattening (b) and upward band bending (c). The corresponding band diagrams are depicted in the insets.}
\label{fig:PC}
\end{figure*}%
Figure \ref{fig:PCcolor}(b) shows the photocurrent dependence as a function of the light injection position and the gate voltage. The absolute value of the photocurrent is maximized near the contacts, whose position is indicated by the yellow lines on the plot. The sign of the photocurrent near the drain is opposite to that near the source. Different amplitudes of $I_{PC}$ around the two contacts, especially at positive gate voltages, are probably due to impurities and surface states at contacts created unintentionally during the fabrication process\cite{kroger_nature_1956,gu_local_2006}. At a gate voltage of $V_S \approx$ -1.25 V the sign inverts at both contacts, showing that the photocurrent can be electrostatically controlled. Further we can see in Figure \ref{fig:PCcolor}(b) that the sensitivity of the channel to illumination for gate voltages above and below $V_S$ is different. In particular, there is an extended region for $V_G > V_S$, where no photocurrent is measured, which indicates that gate-induced doping is sufficiently homogeneous along the channel.\\
The position dependent photocurrent is analyzed in more detail in Figure \ref{fig:PC}, which shows the SPCM line-scans at selected gate voltages (for  V$_{\rm SD}$ = 0 V) in three different regimes $V_G < V_S$, $V_G \approx V_S$ and $V_G > V_S$ (panels (a), (b) and (c) respectively). The electrodes are marked by the yellow regions. The behavior of the photocurrent curves is qualitatively similar, apart from their sign, above and below $V_S$. In either case, we clearly observe two photocurrent peaks in the vicinity of the contacts. The tails of both peaks decrease exponentially as the laser injection position moves away from the respective contact. This behavior is the fingerprint of a device terminated by two Schottky-barriers at the semiconductor-metal interfaces\cite{fu_electrothermal_2011}, where the bands are bent up or down at the semiconductor-metal interface (see insets in Figures \ref{fig:PC}(a) and (c)). The local illumination of the sample excites an electron-hole pair. The minority carrier diffuses towards the nearest contact, where it is extracted due to the built-in electric field, while the majority carrier propagates into the conducting channel and annihilates with a minority carrier created in the other contact\cite{Ahn_Scanning_2005,sze_physics_2006}. The exponential decay is thus given by the diffusion length of the minority carriers. As it was mentioned above, the photocurrent for $V_G > V_S$ decreases faster towards the middle of the sample than in $V_G < V_S$. We note that the position of the current peaks, about 0.5 \mum{} from the contact edges on the exposed sample surface, does not change within our experimental resolution as function of the gate voltage.\\
In the sign-inversion regime, i.e. for $V_G \approx V_S$, Figure \ref{fig:PC}(b), the line-scans are manifestly different from the SPCM profiles found in Figures \ref{fig:PC}(a) and (c). They no longer show two distinguishable peaks and the exponential decay of $I_{PC}$. This gating regime is characterized by band flattening at the sample-contact interface (sketched in the inset of Figure \ref{fig:PC}(b)) and thus the reduction of the electric field near the electrodes. In this situation, the drift currents induced by local potential variation throughout the channel, albeit small, may modify significantly the line-shapes of the SPCM scans\cite{fu_electrothermal_2011}.\\
Apart from the photovoltaic scenario, one should consider a possibility that the photocurrent is partially due to a local heating introduced by the laser and the corresponding thermopower (photothermal effect) \cite{buscema_large_2013}. In the previous studies with a comparable laser power \cite{xu_photo-thermoelectric_2010,buscema_large_2013} the heating of less than 1 K was introduced. The presence of the ionic liquid should make this value even smaller in our case. The Seebeck coefficient of the order of 100 $\mu V/K$ can be deduced directly from the transport characteristic using the Mott relation \cite{xu_photo-thermoelectric_2010,buscema_large_2013}, which agrees with the values directly measured on \ws{} flakes \cite{kim_thermal_2010}. In the OFF state, where the device resistance is above 100 M$\Omega$, the photothermal current should thus not exceed 1 pA, which is more than 3 orders of magnitude smaller than the photocurrents actually observed. Therefore we can ignore this small effect in the further discussion.\\
Now we analyze quantitatively the photocurrent decay in the two band bending regimes (above and below the sign-inversion gate voltage). The minority carrier diffusion length $L_d$ can be extracted by fitting the photocurrent profiles with an exponential decay\cite{fu_electrothermal_2011} as it has been done in experiments with nanowires\cite{balasubramanian_photocurrent_2005} and for 2D materials\cite{graham_scanning_2013}. In order to account for a possible overlap of the photocurrent tails from the two contacts, we use a sum of the two exponential functions
\beq
I_{PC}(x) =& I_{S}\exp\left(-\frac{ x - x_S}{L_{d}}\right) + I_{D}\exp\left(\frac{x - x_D}{L_{d}}\right)
\label{eq:LD}
\eeq%
where the parameters $I_{S,D}$ are the adjustable current amplitudes at the source and drain and $x_{S,D}$ are contact positions, kept fixed in the fitting procedure. The fitting range is hold 1 $\mu$m away from the contact edges, in order to account for the finite size of the laser spot. As it was shown in Ref.\onlinecite{fu_electrothermal_2011} the value of $L_d$ can be correctly obtained in this way if it is smaller than about one-half of the channel length. Thus in our device the diffusion length below 5 \mum{} can be detected. We find satisfactory fits at all gate voltages except in the region close to $V_S$, as expected. As an example, in Figure \ref{fig:difflength}(a) we plot the fitting results for $V_G$ = -0.6 V and -1.7 V representing the cases of up- and downward band bending, respectively. \\%
In Figure \ref{fig:difflength}(b) the diffusion length as a function of the gate voltage (red squares) is displayed. In the studied gating range the value of $L_d$ varies between 0.4 and 2 microns. Importantly, we find that the diffusion length is systematically higher below than above $V_S$. This points to a significant electron-hole asymmetry in \ws{}.\\
The diffusion length is determined by the mobility $\mu$ and the lifetime $\tau$ of the minority charge carriers through the relation $L_d = \sqrt{\frac{k_B T}{e}\mu\tau}$, where $T$ is the temperature, $k_B$ the Boltzmann constant and $e$ the elementary charge\cite{sze_physics_2006}. Therefore the asymmetry of $L_d$ can be due to either one of these parameters or both of them. Recently, a higher mobility was observed for holes than for electrons on similar \ws{} devices\cite{braga_quantitative_2012}, which was tentatively associated to the different density of states in the conduction and valence bands\cite{kuc_influence_2011}. The gate dependence of the diffusion length obtained in the present work matches qualitatively that of the mobility. However an important caveat to keep in mind is that the diffusion length extracted from the SPCM data refers to minority carriers, while electronic transport experiments probe the mobility of majority carriers. Their dependencies on gate and their values are therefore not necessarily the same.\\
For a typical mobility\cite{braga_quantitative_2012} of $\mu$ = 50 cm$^2$V$^{-1}$s$^{-1}$ and a diffusion length of $L_d\approx$ 0.5 \mum{} the carrier lifetime is $\tau$ = 2 ns. This value is in excellent agreement with recent reports on similar experiments on few-layer MoS$_2$\cite{wu_elucidating_2013,choi_high-detectivity_2012}.\\
In addition to the electron-hole asymmetry, we observe in Figure \ref{fig:difflength}(b) a decrease of $L_d$ as the gate voltage moves away from the flat-band voltage $V_S$. This might in part be related to the Shockley-Read-Hall recombination statistics which predicts a decrease of carrier lifetimes with increasing carrier density\cite{sze_physics_2006}.\\
One should mention that the value of band flattening voltage, $V_S$, is very close to the hole threshold voltage, $V_{th,h}$ = -1.4 V, as follows from a comparison between $L_d(V_G)$ and $I_{SD}(V_G)$ that we reproduce on the graph in Figure \ref{fig:difflength}(b) (dashed line). The Ar treatment of the contact area, which can modify the work function of the material, and the presence of the ionic liquid at the semiconductor-metal interface, which affects the properties of the depletion layer\cite{shimotani_electric_2007}, make predictions on the value of $V_S$ difficult. However, this indicates that the microscopic carrier propagation mechanisms for photocurrent and electronic transport are different. Namely, the photocurrent is largely determined by the band bending at the sample-contact interface, while the FET characteristics are given by the charge accumulation at the sample-ionic liquid interface. Irrespective of these considerations, our work clearly shows that both processes are efficiently controlled by ionic liquid gating.\\
In summary, we studied the position dependent generation of photocurrent in \ws{} ambipolar ionic-liquid gated field-effect transistors. We show that the photocurrent can be inverted by gate voltage. The exponential shape of the photocurrent line-scans is observed, which is typical for Schottky-barrier terminated devices except at the gating range where the photocurrent sign is inverted. A strong electron-hole asymmetry is found: the diffusion length of minority electrons is larger than that of minority holes. Our work demonstrates that ionic liquid gating can be combined with scanning photocurrent microscopy. These findings might also be important for future photovoltaic applications of ultra-thin films of transition metal dichalcogenides.%
\begin{acknowledgments}
We acknowledge technical contribution of Romain Bortoli. ABK and AFM gratefully acknowledge financial support by the Swiss National Science Foundation (SNSF), the National Centre of Competence in Research ``Materials with Novel Electronic Properties - MaNEP'', the SNF Sinergia grant on atomically thin transition metal dichalcogenides and the EU Graphene Flagship.
\end{acknowledgments}

\begin{thebibliography}{38}%
\makeatletter
\providecommand \@ifxundefined [1]{%
 \@ifx{#1\undefined}
}%
\providecommand \@ifnum [1]{%
 \ifnum #1\expandafter \@firstoftwo
 \else \expandafter \@secondoftwo
 \fi
}%
\providecommand \@ifx [1]{%
 \ifx #1\expandafter \@firstoftwo
 \else \expandafter \@secondoftwo
 \fi
}%
\providecommand \natexlab [1]{#1}%
\providecommand \enquote  [1]{``#1''}%
\providecommand \bibnamefont  [1]{#1}%
\providecommand \bibfnamefont [1]{#1}%
\providecommand \citenamefont [1]{#1}%
\providecommand \href@noop [0]{\@secondoftwo}%
\providecommand \href [0]{\begingroup \@sanitize@url \@href}%
\providecommand \@href[1]{\@@startlink{#1}\@@href}%
\providecommand \@@href[1]{\endgroup#1\@@endlink}%
\providecommand \@sanitize@url [0]{\catcode `\\12\catcode `\$12\catcode
  `\&12\catcode `\#12\catcode `\^12\catcode `\_12\catcode `\%12\relax}%
\providecommand \@@startlink[1]{}%
\providecommand \@@endlink[0]{}%
\providecommand \url  [0]{\begingroup\@sanitize@url \@url }%
\providecommand \@url [1]{\endgroup\@href {#1}{\urlprefix }}%
\providecommand \urlprefix  [0]{URL }%
\providecommand \Eprint [0]{\href }%
\providecommand \doibase [0]{http://dx.doi.org/}%
\providecommand \selectlanguage [0]{\@gobble}%
\providecommand \bibinfo  [0]{\@secondoftwo}%
\providecommand \bibfield  [0]{\@secondoftwo}%
\providecommand \translation [1]{[#1]}%
\providecommand \BibitemOpen [0]{}%
\providecommand \bibitemStop [0]{}%
\providecommand \bibitemNoStop [0]{.\EOS\space}%
\providecommand \EOS [0]{\spacefactor3000\relax}%
\providecommand \BibitemShut  [1]{\csname bibitem#1\endcsname}%
\let\auto@bib@innerbib\@empty
\bibitem [{\citenamefont {Tributsch}(1979)}]{tributsch_electrochemical_1979}%
  \BibitemOpen
  \bibfield  {author} {\bibinfo {author} {\bibfnamefont {H.}~\bibnamefont
  {Tributsch}},\ }\href {\doibase 10.1016/0165-1633(79)90044-3} {\bibfield
  {journal} {\bibinfo  {journal} {Solar Energy Materials}\ }\textbf {\bibinfo
  {volume} {1}},\ \bibinfo {pages} {257} (\bibinfo {year} {1979})}\BibitemShut
  {NoStop}%
\bibitem [{\citenamefont {Kam}\ and\ \citenamefont
  {Parkinson}(1982)}]{kam_detailed_1982}%
  \BibitemOpen
  \bibfield  {author} {\bibinfo {author} {\bibfnamefont {K.~K.}\ \bibnamefont
  {Kam}}\ and\ \bibinfo {author} {\bibfnamefont {B.~A.}\ \bibnamefont
  {Parkinson}},\ }\href {\doibase 10.1021/j100393a010} {\bibfield  {journal}
  {\bibinfo  {journal} {The Journal of Physical Chemistry}\ }\textbf {\bibinfo
  {volume} {86}},\ \bibinfo {pages} {463} (\bibinfo {year} {1982})}\BibitemShut
  {NoStop}%
\bibitem [{\citenamefont {Decker}, \citenamefont {Scrosati},\ and\
  \citenamefont {Razzini}(1992)}]{decker_photoelectrochemical_1992}%
  \BibitemOpen
  \bibfield  {author} {\bibinfo {author} {\bibfnamefont {F.}~\bibnamefont
  {Decker}}, \bibinfo {author} {\bibfnamefont {B.}~\bibnamefont {Scrosati}}, \
  and\ \bibinfo {author} {\bibfnamefont {G.}~\bibnamefont {Razzini}},\ }in\
  \href {http://link.springer.com/chapter/10.1007/978-94-015-1301-2_3}
  {{\emph {\bibinfo {booktitle} {Photo\-electro\-chemistry and
  Photovoltaics of La\-ye\-red\ Smc}}}},\ \bibinfo {series and number}
  {\bibinfo {series} {Physics and Chemistry of Materials with Low-Dimensional
  Structures}\ No.~\bibinfo {number} {14}},\ \bibinfo {editor} {edited by\
  \bibinfo {editor} {\bibfnamefont {A.}~\bibnamefont {Aruchamy}}}\ (\bibinfo
  {publisher} {Springer Netherlands},\ \bibinfo {year} {1992})\ pp.\ \bibinfo
  {pages} {121--154}\BibitemShut {NoStop}%
\bibitem [{\citenamefont {Gourmelon}\ \emph {et~al.}(1997)\citenamefont
  {Gourmelon}, \citenamefont {Lignier}, \citenamefont {Hadouda}, \citenamefont
  {Couturier}, \citenamefont {Bernède}, \citenamefont {Tedd}, \citenamefont
  {Pouzet},\ and\ \citenamefont {Salardenne}}]{gourmelon_ms2_1997}%
  \BibitemOpen
  \bibfield  {author} {\bibinfo {author} {\bibfnamefont {E.}~\bibnamefont
  {Gourmelon}}, \bibinfo {author} {\bibfnamefont {O.}~\bibnamefont {Lignier}},
  \bibinfo {author} {\bibfnamefont {H.}~\bibnamefont {Hadouda}}, \bibinfo
  {author} {\bibfnamefont {G.}~\bibnamefont {Couturier}}, \bibinfo {author}
  {\bibfnamefont {J.}~\bibnamefont {Bernède}}, \bibinfo {author}
  {\bibfnamefont {J.}~\bibnamefont {Tedd}}, \bibinfo {author} {\bibfnamefont
  {J.}~\bibnamefont {Pouzet}}, \ and\ \bibinfo {author} {\bibfnamefont
  {J.}~\bibnamefont {Salardenne}},\ }\href {\doibase
  10.1016/S0927-0248(96)00096-7} {\bibfield  {journal} {\bibinfo  {journal}
  {Solar Energy Materials and Solar Cells}\ }\textbf {\bibinfo {volume} {46}},\
  \bibinfo {pages} {115} (\bibinfo {year} {1997})}\BibitemShut {NoStop}%
\bibitem [{\citenamefont {Shanmugam}\ \emph {et~al.}(2012)\citenamefont
  {Shanmugam}, \citenamefont {Bansal}, \citenamefont {Durcan},\ and\
  \citenamefont {Yu}}]{shanmugam_schottky-barrier_2012}%
  \BibitemOpen
  \bibfield  {author} {\bibinfo {author} {\bibfnamefont {M.}~\bibnamefont
  {Shanmugam}}, \bibinfo {author} {\bibfnamefont {T.}~\bibnamefont {Bansal}},
  \bibinfo {author} {\bibfnamefont {C.~A.}\ \bibnamefont {Durcan}}, \ and\
  \bibinfo {author} {\bibfnamefont {B.}~\bibnamefont {Yu}},\ }\href {\doibase
  doi:10.1063/1.4773525} {\bibfield  {journal} {\bibinfo  {journal} {Applied
  Physics Letters}\ }\textbf {\bibinfo {volume} {101}},\ \bibinfo {pages}
  {263902} (\bibinfo {year} {2012})}\BibitemShut {NoStop}%
\bibitem [{\citenamefont {Britnell}\ \emph {et~al.}(2013)\citenamefont
  {Britnell}, \citenamefont {Ribeiro}, \citenamefont {Eckmann}, \citenamefont
  {Jalil}, \citenamefont {Belle}, \citenamefont {Mishchenko}, \citenamefont
  {Kim}, \citenamefont {Gorbachev}, \citenamefont {Georgiou}, \citenamefont
  {Morozov}, \citenamefont {Grigorenko}, \citenamefont {Geim}, \citenamefont
  {Casiraghi}, \citenamefont {Neto},\ and\ \citenamefont
  {Novoselov}}]{britnell_strong_2013}%
  \BibitemOpen
  \bibfield  {author} {\bibinfo {author} {\bibfnamefont {L.}~\bibnamefont
  {Britnell}}, \bibinfo {author} {\bibfnamefont {R.~M.}\ \bibnamefont
  {Ribeiro}}, \bibinfo {author} {\bibfnamefont {A.}~\bibnamefont {Eckmann}},
  \bibinfo {author} {\bibfnamefont {R.}~\bibnamefont {Jalil}}, \bibinfo
  {author} {\bibfnamefont {B.~D.}\ \bibnamefont {Belle}}, \bibinfo {author}
  {\bibfnamefont {A.}~\bibnamefont {Mishchenko}}, \bibinfo {author}
  {\bibfnamefont {Y.-J.}\ \bibnamefont {Kim}}, \bibinfo {author} {\bibfnamefont
  {R.~V.}\ \bibnamefont {Gorbachev}}, \bibinfo {author} {\bibfnamefont
  {T.}~\bibnamefont {Georgiou}}, \bibinfo {author} {\bibfnamefont {S.~V.}\
  \bibnamefont {Morozov}}, \bibinfo {author} {\bibfnamefont {A.~N.}\
  \bibnamefont {Grigorenko}}, \bibinfo {author} {\bibfnamefont {A.~K.}\
  \bibnamefont {Geim}}, \bibinfo {author} {\bibfnamefont {C.}~\bibnamefont
  {Casiraghi}}, \bibinfo {author} {\bibfnamefont {A.~H.~C.}\ \bibnamefont
  {Neto}}, \ and\ \bibinfo {author} {\bibfnamefont {K.~S.}\ \bibnamefont
  {Novoselov}},\ }\href {\doibase 10.1126/science.1235547} {\bibfield
  {journal} {\bibinfo  {journal} {Science}\ }\textbf {\bibinfo {volume}
  {340}},\ \bibinfo {pages} {1311} (\bibinfo {year} {2013})},\ \bibinfo {note}
  {{PMID:} 23641062}\BibitemShut {NoStop}%
\bibitem [{\citenamefont {Splendiani}\ \emph {et~al.}(2010)\citenamefont
  {Splendiani}, \citenamefont {Sun}, \citenamefont {Zhang}, \citenamefont {Li},
  \citenamefont {Kim}, \citenamefont {Chim}, \citenamefont {Galli},\ and\
  \citenamefont {Wang}}]{splendiani_emerging_2010}%
  \BibitemOpen
  \bibfield  {author} {\bibinfo {author} {\bibfnamefont {A.}~\bibnamefont
  {Splendiani}}, \bibinfo {author} {\bibfnamefont {L.}~\bibnamefont {Sun}},
  \bibinfo {author} {\bibfnamefont {Y.}~\bibnamefont {Zhang}}, \bibinfo
  {author} {\bibfnamefont {T.}~\bibnamefont {Li}}, \bibinfo {author}
  {\bibfnamefont {J.}~\bibnamefont {Kim}}, \bibinfo {author} {\bibfnamefont
  {C.-Y.}\ \bibnamefont {Chim}}, \bibinfo {author} {\bibfnamefont
  {G.}~\bibnamefont {Galli}}, \ and\ \bibinfo {author} {\bibfnamefont
  {F.}~\bibnamefont {Wang}},\ }\href {\doibase 10.1021/nl903868w} {\bibfield
  {journal} {\bibinfo  {journal} {Nano Letters}\ }\textbf {\bibinfo {volume}
  {10}},\ \bibinfo {pages} {1271} (\bibinfo {year} {2010})}\BibitemShut
  {NoStop}%
\bibitem [{\citenamefont {Mak}\ \emph {et~al.}(2010)\citenamefont {Mak},
  \citenamefont {Lee}, \citenamefont {Hone}, \citenamefont {Shan},\ and\
  \citenamefont {Heinz}}]{mak_atomically_2010}%
  \BibitemOpen
  \bibfield  {author} {\bibinfo {author} {\bibfnamefont {K.~F.}\ \bibnamefont
  {Mak}}, \bibinfo {author} {\bibfnamefont {C.}~\bibnamefont {Lee}}, \bibinfo
  {author} {\bibfnamefont {J.}~\bibnamefont {Hone}}, \bibinfo {author}
  {\bibfnamefont {J.}~\bibnamefont {Shan}}, \ and\ \bibinfo {author}
  {\bibfnamefont {T.~F.}\ \bibnamefont {Heinz}},\ }\href {\doibase
  10.1103/PhysRevLett.105.136805} {\bibfield  {journal} {\bibinfo  {journal}
  {Physical Review Letters}\ }\textbf {\bibinfo {volume} {105}},\ \bibinfo
  {pages} {136805} (\bibinfo {year} {2010})}\BibitemShut {NoStop}%
\bibitem [{\citenamefont {Gutierrez}\ \emph {et~al.}(2012)\citenamefont
  {Gutierrez}, \citenamefont {Perea-Lopez}, \citenamefont {Elias},
  \citenamefont {Berkdemir}, \citenamefont {Wang}, \citenamefont {Lv},
  \citenamefont {Lopez-Urias}, \citenamefont {Crespi}, \citenamefont
  {Terrones},\ and\ \citenamefont {Terrones}}]{gutierrez_extraordinary_2012}%
  \BibitemOpen
  \bibfield  {author} {\bibinfo {author} {\bibfnamefont {H.~R.}\ \bibnamefont
  {Gutierrez}}, \bibinfo {author} {\bibfnamefont {N.}~\bibnamefont
  {Perea-Lopez}}, \bibinfo {author} {\bibfnamefont {A.~L.}\ \bibnamefont
  {Elias}}, \bibinfo {author} {\bibfnamefont {A.}~\bibnamefont {Berkdemir}},
  \bibinfo {author} {\bibfnamefont {B.}~\bibnamefont {Wang}}, \bibinfo {author}
  {\bibfnamefont {R.}~\bibnamefont {Lv}}, \bibinfo {author} {\bibfnamefont
  {F.}~\bibnamefont {Lopez-Urias}}, \bibinfo {author} {\bibfnamefont {V.~H.}\
  \bibnamefont {Crespi}}, \bibinfo {author} {\bibfnamefont {H.}~\bibnamefont
  {Terrones}}, \ and\ \bibinfo {author} {\bibfnamefont {M.}~\bibnamefont
  {Terrones}},\ }\href {\doibase 10.1021/nl3026357} {\bibfield  {journal}
  {\bibinfo  {journal} {Nano Letters}\ } (\bibinfo {year} {2012}),\
  10.1021/nl3026357}\BibitemShut {NoStop}%
\bibitem [{\citenamefont {Zhao}\ \emph {et~al.}(2013)\citenamefont {Zhao},
  \citenamefont {Ghorannevis}, \citenamefont {Chu}, \citenamefont {Toh},
  \citenamefont {Kloc}, \citenamefont {Tan},\ and\ \citenamefont
  {Eda}}]{zhao_evolution_2013}%
  \BibitemOpen
  \bibfield  {author} {\bibinfo {author} {\bibfnamefont {W.}~\bibnamefont
  {Zhao}}, \bibinfo {author} {\bibfnamefont {Z.}~\bibnamefont {Ghorannevis}},
  \bibinfo {author} {\bibfnamefont {L.}~\bibnamefont {Chu}}, \bibinfo {author}
  {\bibfnamefont {M.}~\bibnamefont {Toh}}, \bibinfo {author} {\bibfnamefont
  {C.}~\bibnamefont {Kloc}}, \bibinfo {author} {\bibfnamefont {P.-H.}\
  \bibnamefont {Tan}}, \ and\ \bibinfo {author} {\bibfnamefont
  {G.}~\bibnamefont {Eda}},\ }\href {\doibase 10.1021/nn305275h} {\bibfield
  {journal} {\bibinfo  {journal} {{ACS} Nano}\ }\textbf {\bibinfo {volume}
  {7}},\ \bibinfo {pages} {791} (\bibinfo {year} {2013})}\BibitemShut {NoStop}%
\bibitem [{\citenamefont {Zeng}\ \emph {et~al.}(2013)\citenamefont {Zeng},
  \citenamefont {Liu}, \citenamefont {Dai}, \citenamefont {Yan}, \citenamefont
  {Zhu}, \citenamefont {He}, \citenamefont {Xie}, \citenamefont {Xu},
  \citenamefont {Chen}, \citenamefont {Yao},\ and\ \citenamefont
  {Cui}}]{zeng_optical_2013}%
  \BibitemOpen
  \bibfield  {author} {\bibinfo {author} {\bibfnamefont {H.}~\bibnamefont
  {Zeng}}, \bibinfo {author} {\bibfnamefont {G.-B.}\ \bibnamefont {Liu}},
  \bibinfo {author} {\bibfnamefont {J.}~\bibnamefont {Dai}}, \bibinfo {author}
  {\bibfnamefont {Y.}~\bibnamefont {Yan}}, \bibinfo {author} {\bibfnamefont
  {B.}~\bibnamefont {Zhu}}, \bibinfo {author} {\bibfnamefont {R.}~\bibnamefont
  {He}}, \bibinfo {author} {\bibfnamefont {L.}~\bibnamefont {Xie}}, \bibinfo
  {author} {\bibfnamefont {S.}~\bibnamefont {Xu}}, \bibinfo {author}
  {\bibfnamefont {X.}~\bibnamefont {Chen}}, \bibinfo {author} {\bibfnamefont
  {W.}~\bibnamefont {Yao}}, \ and\ \bibinfo {author} {\bibfnamefont
  {X.}~\bibnamefont {Cui}},\ }\href {\doibase 10.1038/srep01608} {\bibfield
  {journal} {\bibinfo  {journal} {Scientific Reports}\ }\textbf {\bibinfo
  {volume} {3}} (\bibinfo {year} {2013}),\ 10.1038/srep01608}\BibitemShut
  {NoStop}%
\bibitem [{\citenamefont {Radisavljevic}\ \emph {et~al.}(2011)\citenamefont
  {Radisavljevic}, \citenamefont {Radenovic}, \citenamefont {Brivio},
  \citenamefont {Giacometti},\ and\ \citenamefont
  {Kis}}]{radisavljevic_single-layer_2011}%
  \BibitemOpen
  \bibfield  {author} {\bibinfo {author} {\bibfnamefont {B.}~\bibnamefont
  {Radisavljevic}}, \bibinfo {author} {\bibfnamefont {A.}~\bibnamefont
  {Radenovic}}, \bibinfo {author} {\bibfnamefont {J.}~\bibnamefont {Brivio}},
  \bibinfo {author} {\bibfnamefont {V.}~\bibnamefont {Giacometti}}, \ and\
  \bibinfo {author} {\bibfnamefont {A.}~\bibnamefont {Kis}},\ }\href {\doibase
  10.1038/nnano.2010.279} {\bibfield  {journal} {\bibinfo  {journal} {Nature
  Nanotechnology}\ }\textbf {\bibinfo {volume} {6}},\ \bibinfo {pages} {147}
  (\bibinfo {year} {2011})}\BibitemShut {NoStop}%
\bibitem [{\citenamefont {Ahn}, \citenamefont {Dunning},\ and\ \citenamefont
  {Park}(2005)}]{Ahn_Scanning_2005}%
  \BibitemOpen
  \bibfield  {author} {\bibinfo {author} {\bibfnamefont {Y.}~\bibnamefont
  {Ahn}}, \bibinfo {author} {\bibfnamefont {J.}~\bibnamefont {Dunning}}, \ and\
  \bibinfo {author} {\bibfnamefont {J.}~\bibnamefont {Park}},\ }\href {\doibase
  10.1021/nl050631x} {\bibfield  {journal} {\bibinfo  {journal} {Nano Letters}\
  }\textbf {\bibinfo {volume} {5}},\ \bibinfo {pages} {1367} (\bibinfo {year}
  {2005})}\BibitemShut {NoStop}%
\bibitem [{\citenamefont {Yang}\ \emph {et~al.}(2012)\citenamefont {Yang},
  \citenamefont {Li}, \citenamefont {Wu}, \citenamefont {Oh},\ and\
  \citenamefont {Yu}}]{yang_controlled_2012}%
  \BibitemOpen
  \bibfield  {author} {\bibinfo {author} {\bibfnamefont {Y.}~\bibnamefont
  {Yang}}, \bibinfo {author} {\bibfnamefont {J.}~\bibnamefont {Li}}, \bibinfo
  {author} {\bibfnamefont {H.}~\bibnamefont {Wu}}, \bibinfo {author}
  {\bibfnamefont {E.}~\bibnamefont {Oh}}, \ and\ \bibinfo {author}
  {\bibfnamefont {D.}~\bibnamefont {Yu}},\ }\href {\doibase 10.1021/nl303294k}
  {\bibfield  {journal} {\bibinfo  {journal} {Nano Letters}\ }\textbf {\bibinfo
  {volume} {12}},\ \bibinfo {pages} {5890} (\bibinfo {year}
  {2012})}\BibitemShut {NoStop}%
\bibitem [{\citenamefont {Gu}\ \emph {et~al.}(2005)\citenamefont {Gu},
  \citenamefont {Kwak}, \citenamefont {Lensch}, \citenamefont {Allen},
  \citenamefont {Odom},\ and\ \citenamefont {Lauhon}}]{Gu_Near-field_2005}%
  \BibitemOpen
  \bibfield  {author} {\bibinfo {author} {\bibfnamefont {Y.}~\bibnamefont
  {Gu}}, \bibinfo {author} {\bibfnamefont {E.-S.}\ \bibnamefont {Kwak}},
  \bibinfo {author} {\bibfnamefont {J.~L.}\ \bibnamefont {Lensch}}, \bibinfo
  {author} {\bibfnamefont {J.~E.}\ \bibnamefont {Allen}}, \bibinfo {author}
  {\bibfnamefont {T.~W.}\ \bibnamefont {Odom}}, \ and\ \bibinfo {author}
  {\bibfnamefont {L.~J.}\ \bibnamefont {Lauhon}},\ }\href {\doibase
  doi:10.1063/1.1996851} {\bibfield  {journal} {\bibinfo  {journal} {Applied
  Physics Letters}\ }\textbf {\bibinfo {volume} {87}},\ \bibinfo {pages}
  {043111} (\bibinfo {year} {2005})}\BibitemShut {NoStop}%
\bibitem [{\citenamefont {Balasubramanian}\ \emph {et~al.}(2004)\citenamefont
  {Balasubramanian}, \citenamefont {Fan}, \citenamefont {Burghard},
  \citenamefont {Kern}, \citenamefont {Friedrich}, \citenamefont {Wannek},\
  and\ \citenamefont {Mews}}]{Balasubramanian_Photoelectronic_2004}%
  \BibitemOpen
  \bibfield  {author} {\bibinfo {author} {\bibfnamefont {K.}~\bibnamefont
  {Balasubramanian}}, \bibinfo {author} {\bibfnamefont {Y.}~\bibnamefont
  {Fan}}, \bibinfo {author} {\bibfnamefont {M.}~\bibnamefont {Burghard}},
  \bibinfo {author} {\bibfnamefont {K.}~\bibnamefont {Kern}}, \bibinfo {author}
  {\bibfnamefont {M.}~\bibnamefont {Friedrich}}, \bibinfo {author}
  {\bibfnamefont {U.}~\bibnamefont {Wannek}}, \ and\ \bibinfo {author}
  {\bibfnamefont {A.}~\bibnamefont {Mews}},\ }\href {\doibase
  doi:10.1063/1.1688451} {\bibfield  {journal} {\bibinfo  {journal} {Applied
  Physics Letters}\ }\textbf {\bibinfo {volume} {84}},\ \bibinfo {pages} {2400}
  (\bibinfo {year} {2004})}\BibitemShut {NoStop}%
\bibitem [{\citenamefont {Xia}\ \emph {et~al.}(2009)\citenamefont {Xia},
  \citenamefont {Mueller}, \citenamefont {Golizadeh-Mojarad}, \citenamefont
  {Freitag}, \citenamefont {Lin}, \citenamefont {Tsang}, \citenamefont
  {Perebeinos},\ and\ \citenamefont {Avouris}}]{xia_photocurrent_2009}%
  \BibitemOpen
  \bibfield  {author} {\bibinfo {author} {\bibfnamefont {F.}~\bibnamefont
  {Xia}}, \bibinfo {author} {\bibfnamefont {T.}~\bibnamefont {Mueller}},
  \bibinfo {author} {\bibfnamefont {R.}~\bibnamefont {Golizadeh-Mojarad}},
  \bibinfo {author} {\bibfnamefont {M.}~\bibnamefont {Freitag}}, \bibinfo
  {author} {\bibfnamefont {Y.-m.}\ \bibnamefont {Lin}}, \bibinfo {author}
  {\bibfnamefont {J.}~\bibnamefont {Tsang}}, \bibinfo {author} {\bibfnamefont
  {V.}~\bibnamefont {Perebeinos}}, \ and\ \bibinfo {author} {\bibfnamefont
  {P.}~\bibnamefont {Avouris}},\ }\href {\doibase 10.1021/nl8033812} {\bibfield
   {journal} {\bibinfo  {journal} {Nano Letters}\ }\textbf {\bibinfo {volume}
  {9}},\ \bibinfo {pages} {1039} (\bibinfo {year} {2009})}\BibitemShut
  {NoStop}%
\bibitem [{\citenamefont {Buscema}\ \emph {et~al.}(2013)\citenamefont
  {Buscema}, \citenamefont {Barkelid}, \citenamefont {Zwiller}, \citenamefont
  {van~der Zant}, \citenamefont {Steele},\ and\ \citenamefont
  {Castellanos-Gomez}}]{buscema_large_2013}%
  \BibitemOpen
  \bibfield  {author} {\bibinfo {author} {\bibfnamefont {M.}~\bibnamefont
  {Buscema}}, \bibinfo {author} {\bibfnamefont {M.}~\bibnamefont {Barkelid}},
  \bibinfo {author} {\bibfnamefont {V.}~\bibnamefont {Zwiller}}, \bibinfo
  {author} {\bibfnamefont {H.~S.~J.}\ \bibnamefont {van~der Zant}}, \bibinfo
  {author} {\bibfnamefont {G.~A.}\ \bibnamefont {Steele}}, \ and\ \bibinfo
  {author} {\bibfnamefont {A.}~\bibnamefont {Castellanos-Gomez}},\ }\href
  {\doibase 10.1021/nl303321g} {\bibfield  {journal} {\bibinfo  {journal} {Nano
  Letters}\ }\textbf {\bibinfo {volume} {13}},\ \bibinfo {pages} {358}
  (\bibinfo {year} {2013})}\BibitemShut {NoStop}%
\bibitem [{\citenamefont {Wu}\ \emph {et~al.}(2013)\citenamefont {Wu},
  \citenamefont {Jariwala}, \citenamefont {Sangwan}, \citenamefont {Marks},
  \citenamefont {Hersam},\ and\ \citenamefont {Lauhon}}]{wu_elucidating_2013}%
  \BibitemOpen
  \bibfield  {author} {\bibinfo {author} {\bibfnamefont {C.-C.}\ \bibnamefont
  {Wu}}, \bibinfo {author} {\bibfnamefont {D.}~\bibnamefont {Jariwala}},
  \bibinfo {author} {\bibfnamefont {V.~K.}\ \bibnamefont {Sangwan}}, \bibinfo
  {author} {\bibfnamefont {T.~J.}\ \bibnamefont {Marks}}, \bibinfo {author}
  {\bibfnamefont {M.~C.}\ \bibnamefont {Hersam}}, \ and\ \bibinfo {author}
  {\bibfnamefont {L.~J.}\ \bibnamefont {Lauhon}},\ }\href {\doibase
  10.1021/jz401199x} {\bibfield  {journal} {\bibinfo  {journal} {The Journal of
  Physical Chemistry Letters}\ }\textbf {\bibinfo {volume} {4}},\ \bibinfo
  {pages} {2508} (\bibinfo {year} {2013})}\BibitemShut {NoStop}%
\bibitem [{\citenamefont {Podzorov}\ \emph {et~al.}(2004)\citenamefont
  {Podzorov}, \citenamefont {Gershenson}, \citenamefont {Kloc}, \citenamefont
  {Zeis},\ and\ \citenamefont {Bucher}}]{podzorov_high-mobility_2004}%
  \BibitemOpen
  \bibfield  {author} {\bibinfo {author} {\bibfnamefont {V.}~\bibnamefont
  {Podzorov}}, \bibinfo {author} {\bibfnamefont {M.~E.}\ \bibnamefont
  {Gershenson}}, \bibinfo {author} {\bibfnamefont {C.}~\bibnamefont {Kloc}},
  \bibinfo {author} {\bibfnamefont {R.}~\bibnamefont {Zeis}}, \ and\ \bibinfo
  {author} {\bibfnamefont {E.}~\bibnamefont {Bucher}},\ }\href {\doibase
  doi:10.1063/1.1723695} {\bibfield  {journal} {\bibinfo  {journal} {Applied
  Physics Letters}\ }\textbf {\bibinfo {volume} {84}},\ \bibinfo {pages} {3301}
  (\bibinfo {year} {2004})}\BibitemShut {NoStop}%
\bibitem [{\citenamefont {Sik~Hwang}\ \emph {et~al.}(2012)\citenamefont
  {Sik~Hwang}, \citenamefont {Remskar}, \citenamefont {Yan}, \citenamefont
  {Protasenko}, \citenamefont {Tahy}, \citenamefont {Doo~Chae}, \citenamefont
  {Zhao}, \citenamefont {Konar}, \citenamefont {(Grace)~Xing}, \citenamefont
  {Seabaugh},\ and\ \citenamefont {Jena}}]{sik_hwang_transistors_2012}%
  \BibitemOpen
  \bibfield  {author} {\bibinfo {author} {\bibfnamefont {W.}~\bibnamefont
  {Sik~Hwang}}, \bibinfo {author} {\bibfnamefont {M.}~\bibnamefont {Remskar}},
  \bibinfo {author} {\bibfnamefont {R.}~\bibnamefont {Yan}}, \bibinfo {author}
  {\bibfnamefont {V.}~\bibnamefont {Protasenko}}, \bibinfo {author}
  {\bibfnamefont {K.}~\bibnamefont {Tahy}}, \bibinfo {author} {\bibfnamefont
  {S.}~\bibnamefont {Doo~Chae}}, \bibinfo {author} {\bibfnamefont
  {P.}~\bibnamefont {Zhao}}, \bibinfo {author} {\bibfnamefont {A.}~\bibnamefont
  {Konar}}, \bibinfo {author} {\bibfnamefont {H.}~\bibnamefont {(Grace)~Xing}},
  \bibinfo {author} {\bibfnamefont {A.}~\bibnamefont {Seabaugh}}, \ and\
  \bibinfo {author} {\bibfnamefont {D.}~\bibnamefont {Jena}},\ }\href {\doibase
  doi:10.1063/1.4732522} {\bibfield  {journal} {\bibinfo  {journal} {Applied
  Physics Letters}\ }\textbf {\bibinfo {volume} {101}},\ \bibinfo {pages}
  {013107} (\bibinfo {year} {2012})}\BibitemShut {NoStop}%
\bibitem [{\citenamefont {Panzer}, \citenamefont {Newman},\ and\ \citenamefont
  {Frisbie}(2005)}]{panzer_low-voltage_2005}%
  \BibitemOpen
  \bibfield  {author} {\bibinfo {author} {\bibfnamefont {M.~J.}\ \bibnamefont
  {Panzer}}, \bibinfo {author} {\bibfnamefont {C.~R.}\ \bibnamefont {Newman}},
  \ and\ \bibinfo {author} {\bibfnamefont {C.~D.}\ \bibnamefont {Frisbie}},\
  }\href {\doibase doi:10.1063/1.1880434} {\bibfield  {journal} {\bibinfo
  {journal} {Applied Physics Letters}\ }\textbf {\bibinfo {volume} {86}},\
  \bibinfo {pages} {103503} (\bibinfo {year} {2005})}\BibitemShut {NoStop}%
\bibitem [{\citenamefont {Shimotani}\ \emph {et~al.}(2006)\citenamefont
  {Shimotani}, \citenamefont {Asanuma}, \citenamefont {Takeya},\ and\
  \citenamefont {Iwasa}}]{shimotani_electrolyte-gated_2006}%
  \BibitemOpen
  \bibfield  {author} {\bibinfo {author} {\bibfnamefont {H.}~\bibnamefont
  {Shimotani}}, \bibinfo {author} {\bibfnamefont {H.}~\bibnamefont {Asanuma}},
  \bibinfo {author} {\bibfnamefont {J.}~\bibnamefont {Takeya}}, \ and\ \bibinfo
  {author} {\bibfnamefont {Y.}~\bibnamefont {Iwasa}},\ }\href {\doibase
  doi:10.1063/1.2387884} {\bibfield  {journal} {\bibinfo  {journal} {Applied
  Physics Letters}\ }\textbf {\bibinfo {volume} {89}},\ \bibinfo {pages}
  {203501} (\bibinfo {year} {2006})}\BibitemShut {NoStop}%
\bibitem [{\citenamefont {Ueno}\ \emph {et~al.}(2008)\citenamefont {Ueno},
  \citenamefont {Nakamura}, \citenamefont {Shimotani}, \citenamefont {Ohtomo},
  \citenamefont {Kimura}, \citenamefont {Nojima}, \citenamefont {Aoki},
  \citenamefont {Iwasa},\ and\ \citenamefont
  {Kawasaki}}]{ueno_electric-field-induced_2008}%
  \BibitemOpen
  \bibfield  {author} {\bibinfo {author} {\bibfnamefont {K.}~\bibnamefont
  {Ueno}}, \bibinfo {author} {\bibfnamefont {S.}~\bibnamefont {Nakamura}},
  \bibinfo {author} {\bibfnamefont {H.}~\bibnamefont {Shimotani}}, \bibinfo
  {author} {\bibfnamefont {A.}~\bibnamefont {Ohtomo}}, \bibinfo {author}
  {\bibfnamefont {N.}~\bibnamefont {Kimura}}, \bibinfo {author} {\bibfnamefont
  {T.}~\bibnamefont {Nojima}}, \bibinfo {author} {\bibfnamefont
  {H.}~\bibnamefont {Aoki}}, \bibinfo {author} {\bibfnamefont {Y.}~\bibnamefont
  {Iwasa}}, \ and\ \bibinfo {author} {\bibfnamefont {M.}~\bibnamefont
  {Kawasaki}},\ }\href {\doibase 10.1038/nmat2298} {\bibfield  {journal}
  {\bibinfo  {journal} {Nature Materials}\ }\textbf {\bibinfo {volume} {7}},\
  \bibinfo {pages} {855} (\bibinfo {year} {2008})}\BibitemShut {NoStop}%
\bibitem [{\citenamefont {Braga}\ \emph {et~al.}(2012)\citenamefont {Braga},
  \citenamefont {Gutiérrez~Lezama}, \citenamefont {Berger},\ and\
  \citenamefont {Morpurgo}}]{braga_quantitative_2012}%
  \BibitemOpen
  \bibfield  {author} {\bibinfo {author} {\bibfnamefont {D.}~\bibnamefont
  {Braga}}, \bibinfo {author} {\bibfnamefont {I.}~\bibnamefont
  {Gutiérrez~Lezama}}, \bibinfo {author} {\bibfnamefont {H.}~\bibnamefont
  {Berger}}, \ and\ \bibinfo {author} {\bibfnamefont {A.~F.}\ \bibnamefont
  {Morpurgo}},\ }\href {\doibase 10.1021/nl302389d} {\bibfield  {journal}
  {\bibinfo  {journal} {Nano Letters}\ }\textbf {\bibinfo {volume} {12}},\
  \bibinfo {pages} {5218} (\bibinfo {year} {2012})}\BibitemShut {NoStop}%
\bibitem [{\citenamefont {Ye}\ \emph {et~al.}(2012)\citenamefont {Ye},
  \citenamefont {Zhang}, \citenamefont {Akashi}, \citenamefont {Bahramy},
  \citenamefont {Arita},\ and\ \citenamefont
  {Iwasa}}]{ye_superconducting_2012}%
  \BibitemOpen
  \bibfield  {author} {\bibinfo {author} {\bibfnamefont {J.~T.}\ \bibnamefont
  {Ye}}, \bibinfo {author} {\bibfnamefont {Y.~J.}\ \bibnamefont {Zhang}},
  \bibinfo {author} {\bibfnamefont {R.}~\bibnamefont {Akashi}}, \bibinfo
  {author} {\bibfnamefont {M.~S.}\ \bibnamefont {Bahramy}}, \bibinfo {author}
  {\bibfnamefont {R.}~\bibnamefont {Arita}}, \ and\ \bibinfo {author}
  {\bibfnamefont {Y.}~\bibnamefont {Iwasa}},\ }\href {\doibase
  10.1126/science.1228006} {\bibfield  {journal} {\bibinfo  {journal}
  {Science}\ }\textbf {\bibinfo {volume} {338}},\ \bibinfo {pages} {1193}
  (\bibinfo {year} {2012})},\ \bibinfo {note} {{PMID:} 23197529}\BibitemShut
  {NoStop}%
\bibitem [{\citenamefont {Zhang}\ \emph {et~al.}(2012)\citenamefont {Zhang},
  \citenamefont {Ye}, \citenamefont {Matsuhashi},\ and\ \citenamefont
  {Iwasa}}]{zhang_ambipolar_2012}%
  \BibitemOpen
  \bibfield  {author} {\bibinfo {author} {\bibfnamefont {Y.}~\bibnamefont
  {Zhang}}, \bibinfo {author} {\bibfnamefont {J.}~\bibnamefont {Ye}}, \bibinfo
  {author} {\bibfnamefont {Y.}~\bibnamefont {Matsuhashi}}, \ and\ \bibinfo
  {author} {\bibfnamefont {Y.}~\bibnamefont {Iwasa}},\ }\href {\doibase
  10.1021/nl2021575} {\bibfield  {journal} {\bibinfo  {journal} {Nano Letters}\
  }\textbf {\bibinfo {volume} {12}},\ \bibinfo {pages} {1136} (\bibinfo {year}
  {2012})}\BibitemShut {NoStop}%
\bibitem [{\citenamefont {Kröger}, \citenamefont {Diemer},\ and\ \citenamefont
  {Klasens}(1956)}]{kroger_nature_1956}%
  \BibitemOpen
  \bibfield  {author} {\bibinfo {author} {\bibfnamefont {F.~A.}\ \bibnamefont
  {Kröger}}, \bibinfo {author} {\bibfnamefont {G.}~\bibnamefont {Diemer}}, \
  and\ \bibinfo {author} {\bibfnamefont {H.~A.}\ \bibnamefont {Klasens}},\
  }\href {\doibase 10.1103/PhysRev.103.279} {\bibfield  {journal} {\bibinfo
  {journal} {Physical Review}\ }\textbf {\bibinfo {volume} {103}},\ \bibinfo
  {pages} {279} (\bibinfo {year} {1956})}\BibitemShut {NoStop}%
\bibitem [{\citenamefont {Gu}\ \emph {et~al.}(2006)\citenamefont {Gu},
  \citenamefont {Romankiewicz}, \citenamefont {David}, \citenamefont {Lensch},
  \citenamefont {Lauhon}, \citenamefont {Kwak},\ and\ \citenamefont
  {Odom}}]{gu_local_2006}%
  \BibitemOpen
  \bibfield  {author} {\bibinfo {author} {\bibfnamefont {Y.}~\bibnamefont
  {Gu}}, \bibinfo {author} {\bibfnamefont {J.~P.}\ \bibnamefont
  {Romankiewicz}}, \bibinfo {author} {\bibfnamefont {J.~K.}\ \bibnamefont
  {David}}, \bibinfo {author} {\bibfnamefont {J.~L.}\ \bibnamefont {Lensch}},
  \bibinfo {author} {\bibfnamefont {L.~J.}\ \bibnamefont {Lauhon}}, \bibinfo
  {author} {\bibfnamefont {E.-S.}\ \bibnamefont {Kwak}}, \ and\ \bibinfo
  {author} {\bibfnamefont {T.~W.}\ \bibnamefont {Odom}},\ }\href {\doibase
  10.1116/1.2216717} {\bibfield  {journal} {\bibinfo  {journal} {Journal of
  Vacuum Science \& Technology B}\ }\textbf {\bibinfo {volume} {24}},\ \bibinfo
  {pages} {2172} (\bibinfo {year} {2006})}\BibitemShut {NoStop}%
\bibitem [{\citenamefont {Fu}\ \emph {et~al.}(2011)\citenamefont {Fu},
  \citenamefont {Zou}, \citenamefont {Wang}, \citenamefont {Zhang},
  \citenamefont {Yu},\ and\ \citenamefont {Wu}}]{fu_electrothermal_2011}%
  \BibitemOpen
  \bibfield  {author} {\bibinfo {author} {\bibfnamefont {D.}~\bibnamefont
  {Fu}}, \bibinfo {author} {\bibfnamefont {J.}~\bibnamefont {Zou}}, \bibinfo
  {author} {\bibfnamefont {K.}~\bibnamefont {Wang}}, \bibinfo {author}
  {\bibfnamefont {R.}~\bibnamefont {Zhang}}, \bibinfo {author} {\bibfnamefont
  {D.}~\bibnamefont {Yu}}, \ and\ \bibinfo {author} {\bibfnamefont
  {J.}~\bibnamefont {Wu}},\ }\href {\doibase 10.1021/nl2018806} {\bibfield
  {journal} {\bibinfo  {journal} {Nano Letters}\ }\textbf {\bibinfo {volume}
  {11}},\ \bibinfo {pages} {3809} (\bibinfo {year} {2011})}\BibitemShut
  {NoStop}%
\bibitem [{\citenamefont {Sze}\ and\ \citenamefont
  {Ng}(2006)}]{sze_physics_2006}%
  \BibitemOpen
  \bibfield  {author} {\bibinfo {author} {\bibfnamefont {S.~M.}\ \bibnamefont
  {Sze}}\ and\ \bibinfo {author} {\bibfnamefont {K.~K.}\ \bibnamefont {Ng}},\
  }\href
  {file:///austausch/buecher/Simon_M._Sze,_Kwok_K._Ng_Physics_of_Semiconductor_Devices____2006.pdf}
  {\emph {\bibinfo {title} {Physics of Semiconductor Devices}}}\ (\bibinfo
  {publisher} {John Wiley \& Sons},\ \bibinfo {year} {2006})\BibitemShut
  {NoStop}%
\bibitem [{\citenamefont {Xu}\ \emph {et~al.}(2010)\citenamefont {Xu},
  \citenamefont {Gabor}, \citenamefont {Alden}, \citenamefont {van~der Zande},\
  and\ \citenamefont {{McEuen}}}]{xu_photo-thermoelectric_2010}%
  \BibitemOpen
  \bibfield  {author} {\bibinfo {author} {\bibfnamefont {X.}~\bibnamefont
  {Xu}}, \bibinfo {author} {\bibfnamefont {N.~M.}\ \bibnamefont {Gabor}},
  \bibinfo {author} {\bibfnamefont {J.~S.}\ \bibnamefont {Alden}}, \bibinfo
  {author} {\bibfnamefont {A.~M.}\ \bibnamefont {van~der Zande}}, \ and\
  \bibinfo {author} {\bibfnamefont {P.~L.}\ \bibnamefont {{McEuen}}},\ }\href
  {\doibase 10.1021/nl903451y} {\bibfield  {journal} {\bibinfo  {journal} {Nano
  Letters}\ }\textbf {\bibinfo {volume} {10}},\ \bibinfo {pages} {562}
  (\bibinfo {year} {2010})}\BibitemShut {NoStop}%
\bibitem [{\citenamefont {Kim}\ \emph {et~al.}(2010)\citenamefont {Kim},
  \citenamefont {Choi}, \citenamefont {Seo},\ and\ \citenamefont
  {Cho}}]{kim_thermal_2010}%
  \BibitemOpen
  \bibfield  {author} {\bibinfo {author} {\bibfnamefont {J.-Y.}\ \bibnamefont
  {Kim}}, \bibinfo {author} {\bibfnamefont {S.-M.}\ \bibnamefont {Choi}},
  \bibinfo {author} {\bibfnamefont {W.-S.}\ \bibnamefont {Seo}}, \ and\
  \bibinfo {author} {\bibfnamefont {W.-S.}\ \bibnamefont {Cho}},\ }\href
  {\doibase 10.5012/bkcs.2010.31.11.3225} {\bibfield  {journal} {\bibinfo
  {journal} {Bulletin of the Korean Chemical Society}\ }\textbf {\bibinfo
  {volume} {31}},\ \bibinfo {pages} {3225} (\bibinfo {year}
  {2010})}\BibitemShut {NoStop}%
\bibitem [{\citenamefont {Balasubramanian}\ \emph {et~al.}(2005)\citenamefont
  {Balasubramanian}, \citenamefont {Burghard}, \citenamefont {Kern},
  \citenamefont {Scolari},\ and\ \citenamefont
  {Mews}}]{balasubramanian_photocurrent_2005}%
  \BibitemOpen
  \bibfield  {author} {\bibinfo {author} {\bibfnamefont {K.}~\bibnamefont
  {Balasubramanian}}, \bibinfo {author} {\bibfnamefont {M.}~\bibnamefont
  {Burghard}}, \bibinfo {author} {\bibfnamefont {K.}~\bibnamefont {Kern}},
  \bibinfo {author} {\bibfnamefont {M.}~\bibnamefont {Scolari}}, \ and\
  \bibinfo {author} {\bibfnamefont {A.}~\bibnamefont {Mews}},\ }\href {\doibase
  10.1021/nl050053k} {\bibfield  {journal} {\bibinfo  {journal} {Nano Letters}\
  }\textbf {\bibinfo {volume} {5}},\ \bibinfo {pages} {507} (\bibinfo {year}
  {2005})}\BibitemShut {NoStop}%
\bibitem [{\citenamefont {Graham}\ and\ \citenamefont
  {Yu}(2013)}]{graham_scanning_2013}%
  \BibitemOpen
  \bibfield  {author} {\bibinfo {author} {\bibfnamefont {R.}~\bibnamefont
  {Graham}}\ and\ \bibinfo {author} {\bibfnamefont {D.}~\bibnamefont {Yu}},\
  }\href {\doibase 10.1142/S0217984913300184} {\bibfield  {journal} {\bibinfo
  {journal} {Mod. Phys. Lett. B}\ }\textbf {\bibinfo {volume} {27}},\
  \bibinfo {pages} {1330018} (\bibinfo {year} {2013})}\BibitemShut {NoStop}%
\bibitem [{\citenamefont {Kuc}, \citenamefont {Zibouche},\ and\ \citenamefont
  {Heine}(2011)}]{kuc_influence_2011}%
  \BibitemOpen
  \bibfield  {author} {\bibinfo {author} {\bibfnamefont {A.}~\bibnamefont
  {Kuc}}, \bibinfo {author} {\bibfnamefont {N.}~\bibnamefont {Zibouche}}, \
  and\ \bibinfo {author} {\bibfnamefont {T.}~\bibnamefont {Heine}},\ }\href
  {\doibase 10.1103/PhysRevB.83.245213} {\bibfield  {journal} {\bibinfo
  {journal} {Physical Review B}\ }\textbf {\bibinfo {volume} {83}},\ \bibinfo
  {pages} {245213} (\bibinfo {year} {2011})}\BibitemShut {NoStop}%
\bibitem [{\citenamefont {Choi}\ \emph {et~al.}(2012)\citenamefont {Choi},
  \citenamefont {Cho}, \citenamefont {Konar}, \citenamefont {Lee},
  \citenamefont {Cha}, \citenamefont {Hong}, \citenamefont {Kim}, \citenamefont
  {Kim}, \citenamefont {Jena}, \citenamefont {Joo},\ and\ \citenamefont
  {Kim}}]{choi_high-detectivity_2012}%
  \BibitemOpen
  \bibfield  {author} {\bibinfo {author} {\bibfnamefont {W.}~\bibnamefont
  {Choi}}, \bibinfo {author} {\bibfnamefont {M.~Y.}\ \bibnamefont {Cho}},
  \bibinfo {author} {\bibfnamefont {A.}~\bibnamefont {Konar}}, \bibinfo
  {author} {\bibfnamefont {J.~H.}\ \bibnamefont {Lee}}, \bibinfo {author}
  {\bibfnamefont {G.-B.}\ \bibnamefont {Cha}}, \bibinfo {author} {\bibfnamefont
  {S.~C.}\ \bibnamefont {Hong}}, \bibinfo {author} {\bibfnamefont
  {S.}~\bibnamefont {Kim}}, \bibinfo {author} {\bibfnamefont {J.}~\bibnamefont
  {Kim}}, \bibinfo {author} {\bibfnamefont {D.}~\bibnamefont {Jena}}, \bibinfo
  {author} {\bibfnamefont {J.}~\bibnamefont {Joo}}, \ and\ \bibinfo {author}
  {\bibfnamefont {S.}~\bibnamefont {Kim}},\ }\href {\doibase
  10.1002/adma.201201909} {\bibfield  {journal} {\bibinfo  {journal} {Advanced
  Materials}\ }\textbf {\bibinfo {volume} {24}},\ \bibinfo {pages}
  {5832–5836} (\bibinfo {year} {2012})}\BibitemShut {NoStop}%
\bibitem [{\citenamefont {Shimotani}, \citenamefont {Asanuma},\ and\
  \citenamefont {Iwasa}(2007)}]{shimotani_electric_2007}%
  \BibitemOpen
  \bibfield  {author} {\bibinfo {author} {\bibfnamefont {H.}~\bibnamefont
  {Shimotani}}, \bibinfo {author} {\bibfnamefont {H.}~\bibnamefont {Asanuma}},
  \ and\ \bibinfo {author} {\bibfnamefont {Y.}~\bibnamefont {Iwasa}},\ }\href
  {\doibase 10.7567/JJAP.46.3613} {\bibfield  {journal} {\bibinfo  {journal}
  {Japanese Journal of Applied Physics}\ }\textbf {\bibinfo {volume} {46}},\
  \bibinfo {pages} {3613} (\bibinfo {year} {2007})}\BibitemShut {NoStop}%
\end{thebibliography}
%

\end{document}